\documentstyle[12pt] {article}
\def\caja{\mathsurround=0pt}
\def\eqalign#1{\,\vcenter{\openup1\jot \caja
        \ialign{\strut \hfil$\displaystyle{##}$&$
        \displaystyle{{}##}$\hfil\crcr#1\crcr}}\,}
\begin{document}
\begin{flushright}
GUTPA/91/11-1\\
\end{flushright}
\vskip .2 cm
\begin{center}
{\bf \Large THE GRAND CANONICAL PARTITION FUNCTION\\ \vskip .1in
OF A 2-DIMENSIONAL HUBBARD MODEL}
\vskip 1.cm
{\Large I.M. Barbour and E.G. Klepfish}
\vskip .1in

Dept. of Physics and Astronomy\\
University of Glasgow\\
Glasgow G12 8QQ, U.K.
\end{center}
\baselineskip 28 pt
\begin{center}
\bf Abstract
\end{center}
\noindent
We present a new technique for
a numerical analysis of the phase structure
of the 2D Hubbard model as a function of the hole chemical potential.
The grand canonical partition function for the model
is obtained via Monte Carlo simulations. The dependence of
the hole occupation number on the
chemical potential and the temperature
is evaluated. These calculations,
together with a study of the Yang-Lee
zeros of the grand canonical partition function,
show evidence of a phase transition at zero
temperature and particle density below half-filling.
The binding energy of a pair of holes is calculated in the
low temperature regime
and the possibility for pairing is explored.
\newpage
\bigskip
\noindent{\bf 1. Introduction}
\bigskip\newline\noindent
For many years
the Hubbard model, and other related
systems with a finite density of electrons, has attracted
much attention in the field of numerical
simulations [\ref{Hirsch}],[\ref{Blankenbecler}].
The main interest in these simulations
arose from the suggested relation between the planar Hubbard
model and high T$_c$ superconductivity (HTSC) [\ref{Dagotto}],
[\ref{White}],[\ref{bookH}].
Since the model is
non-relativistic, its analysis avoids
some of the problems associated with relativistic
fermions, such as fermion doubling
[\ref{Blankenbecler}]. However, the inherent
difficulties of simulating fermions
at finite density remain:
integration over the fermionic degrees of freedom
leads to a non-positive integration
measure in the path integrals. This arises from the
determinant of the fermion
matrix being non-positive definite and any importance sampling based
on a partition function proportional to this determinant loses
its normal meaning.
\par In the Hubbard model, this problem manifests itself
via the so-called sign problem of the partition function.
The partition function of the half-filled Hubbard model (one electron
per lattice site) is always positive, being a product
of a determinant of a matrix and its hermitian conjugate. However,
if one introduces a finite {\it real}
chemical potential for the impurities (holes), then this
positivity is lost, and configurations with real determinants, but with
negative signs can occur. Unfortunately, the finite
density of holes is the case of greatest physical interest, as
the impurities play an essential role in the superconducting
transition [\ref{Anderson}].
\par In this paper we propose an analysis of the Hubbard model
which treats the fermion dynamics in a rigorous manner.
This method is similar to one applied to the chiral phase
transition at finite density QCD, and is based on a study of the
zeros of the grand canonical partition function in the
complex fugacity ($e^{\beta \mu}$) plane
[\ref{YangLee}],
[\ref{BarbourZabeur}].
Here we simulate the configurations of the Hubbard-Stratonovich
(HS)
fields (given by Ising variables) at half-filling, as well as at
finite doping fraction
and expand
the grand canonical
partition function (GCPF)
as a polynomial in $e^{\beta\mu}$ (or equivalently in
$e^{\beta\mu}+e^{-\beta\mu}$)[\ref{Dagotto}].
The coefficients of this polynomial
are averaged over an equilibrated
ensemble of configurations. The distribution of the zeros
of this polynomial in the neighbourhood of the physical region,
$\mu$ real,  can indicate a phase
transition. In particular, their scaling behaviour with respect
to the lattice size can indicate if real zeros can occur in the
infinite lattice limit. Such zeros would correspond to
divergences within the theory
at the corresponding
values of the fugacity.
\par The simulation is performed on a 2D
spatial square lattice with the third euclidean time dimension
corresponding to the inverse temperature $\beta$. The updating procedure
is the one described by White {\it et al.} [\ref{White}].
\par As we show below,
the evaluation of the partition function as an explicit
polynomial in the fugacity variable permits analysis
of the superconducting properties of the model for various
values of the hole density.
\par In Section 2 we describe the construction of the
partition function of the Hubbard model as a polynomial in the fugacity
variable.
Section 3 summarizes the simulation
procedure and the the measurements, and in
Section 4 we present the preliminary numerical results for the critical
value of the chemical potential.
\bigskip
\newline\noindent{\bf 2. Finite-temperature partition function}
\bigskip\noindent\newline
The original Hubbard hamiltonian is
given by:
\begin{equation}\label{Hubbard}
H=-t\sum_{i,j,\sigma}c_{i,\sigma}^\dagger c_{j,\sigma}+
U\sum_{i}(n_{i+}-{1\over 2})(n_{i-}-{1\over 2})-\mu\sum_{i}(n_{i+}+
n_{i-})
\end{equation}
where the $i,j$ denote the nearest neighbour spatial
lattice sites, $\sigma$ is
the spin degree of
freedom and $n_{i\sigma}$ is the electron number operator
$c_{i\sigma}^\dagger c_{i\sigma}$. The constants $t$ and $U$ correspond
to the hopping parameter and the on-site Coulomb repulsion respectively.
The chemical potential $\mu$ is introduced such that $\mu=0$
corresponds to half-filling.
\par The finite temperature grand canonical
partition function (GCPF) is given by:
\begin{equation}\label{partition}
Z=Tr(e^{-\beta H})
\end{equation}
where $\beta$ is the inverse temperature.
\par The finite temperature is represented on the lattice
by extending the spatial
lattice in the imaginary-time direction and relating the inverse
temperature $\beta$ to the length of the time dimension $n_\tau$
by
$\beta=n_\tau dt$, where $dt$ is the length of the time step. Following
Hirsch [\ref{Hirsch}] and White {\it et al.} [\ref{White}] we rewrite
the partition function as:
\begin{equation}\label{partone}
Z=Tr(e^{-dtV}e^{-dtK}e^{dt \mu})^{n_\tau}
\end{equation}
where $K$ corresponds to the nearest neighbour
hopping term in the Hubbard
hamiltonian (\ref{Hubbard}) and
$V$ to the onsite interaction
including quartic products of fermion fields.
This decomposition, based on the Trotter
formula [\ref{Suzuki}],
introduces a systematic error proportional
to $dt^2$. The quartic interaction can be rewritten in terms of
Ising fields $s_{i,l}$ using the discrete Hubbard-Stratonovich
transformation [\ref{Hirsch}]:
\begin{equation}\label{Vnew}
e^{-dtV}=e^{-{{dtU}\over 4}}{1\over 2}\sum_{s_{i,l}=\pm 1}
e^{-dts_{i,l}\lambda(n_{i+}-n_{i-})}
\end{equation}
where $i,l$ is the space-time index of a lattice site and
the coupling $\lambda$ is related to the original on-site
repulsion constant by:
\begin{equation}
\cosh{(dt \lambda)}=\exp{({{dt U}\over2})}.
\end{equation}
\par This linearization of the interaction enables one to integrate out
the fermionic degrees of freedom and the resulting partition
function is written as an ensemble average of
a product of two determinants:
\begin{equation}\label{parttwo}
Z=
\sum_{s_{i,l}=\pm 1} \tilde z =
\sum_{s_{i,l}=\pm 1} det(M^+)det(M^-)
\end{equation}
such that
\begin{equation}\label{defM}
M^\pm = (I+P^\pm)=(I+\prod_{l=1}^{n_\tau}B_l^{\pm})
\end{equation}
where
the matrices $B_l^\pm $ are
defined as
\begin{equation}\label{Bls}
B_l^\pm=e^{-(\pm dt V)}e^{-dtK}e^{dt \mu}
\end{equation}
with $V_{ij}=
\delta_{ij}s_{i,l}$ and $K$
the matrix connecting
nearest-neighbours sites with the hopping parameter $t=1$.
The matrices in (\ref{defM}) and (\ref{Bls}) are of size
$(n_xn_y)\times(n_xn_y)$,
corresponding to the spatial size of the lattice. However,
$det(M^\pm)$ can be represented as a determinant of an $(n_xn_yn_\tau)$
square matrix of the form
[\ref{Blankenbecler}]:
\begin{equation}\label{bigmatrix}
{I+A^\pm=\left(\matrix{I&0& ...&B_{1}^\pm\cr
-B_2^\pm&I&...&0\cr
0&...&...&... \cr
...&...&...&0 \cr
0&...&-B_{n_\tau}^\pm&I\cr}\right)},
\end{equation}
a fact which is exploited below in the evaluation of the
partition function $Z$.
\par The expectation value of a physical
observable at chemical potential
$\mu$,
$<O>_\mu$, is given
by:
\begin{equation}\label{observa}
<O>_\mu = {{\int O \tilde z(\mu)}\over {\int \tilde z(\mu)}}
\end{equation}
where the sum over the configurations of Ising fields
is denoted by an integral
Since $\tilde z(\mu)$ is not positive definite for Re$(\mu) \neq 0$
we weight the ensemble of configurations by the absolute
value of $\tilde z(\mu)$ at some $\mu = \mu_0$. Thus
\begin{equation}\label{observb}
<O>_\mu =
{{\int{{O \tilde z(\mu)}\over{|\tilde z(\mu_0)|}}
|\tilde z(\mu_0)|}\over
{\int {{\tilde z(\mu)}\over {|\tilde z(\mu_0)|}}
|\tilde z(\mu_0)|}}
={{<{{O\tilde z(\mu)}\over {|\tilde z(\mu_0)|}}>_{\mu_0}}\over
{<{{\tilde z(\mu)}
\over {|\tilde z(\mu_0)|}}>_{\mu_0}}}
\end{equation}
The partition function $Z(\mu)$ is given by
\begin{equation}\label{newzz}
Z(\mu)\propto<{{\tilde z(\mu)}\over {|\tilde z(\mu_0)|}}>_{\mu_0}.
\end{equation}
 The normalization of the GCPF
is irrelevant as can be seen from eq.(\ref{observb}).
\par The particle-hole transformation
[\ref{Hirsch}],[\ref{SugarLat}]
\begin{equation}\label{particlehole}
d_{i\sigma}=(-1)^ic_{i,\sigma}^\dagger
\end{equation}
is equivalent to the unitary transformation
\begin{equation}\label{trans}
{B_l^+\rightarrow
\left(\matrix{i&0&...&0\cr
0&-i&0&...\cr
0&0&i&...\cr
0&...&0&-i\cr
}\right) B_l^+
\left(\matrix{-i&0&...&0\cr
0&i&0&...\cr
0&0&-i&...\cr
0&...&0&i\cr
}\right)}
\end{equation}
which reverses the sign of the hopping term $K$.
Hence on an even sized lattice
the determinant of
$e^{dtK}$ is 1.
Applying this transformation
to the statistical weight gives:
\begin{equation}\label{partnew}
\tilde z=det(Ie^{-\beta \mu}+P^-_{|_{\mu=0}})
det(Ie^{\beta \mu}+(P^-_{|_{\mu=0}})^\dagger)
e^{\mu n_xn_y\beta}e^{\lambda \sum_{i,l}s_{i,l}}.
\end{equation}
\par Equation (\ref{partnew}) suggests two different ways of
expansion of the partition function. The first one based on
\begin{equation}\label{firstz}
\eqalign{
\tilde z(\mu)=&
{{\prod_{\lambda_i}
((e^{\mu\beta}+e^{-\mu\beta})+(\lambda_i+{1\over {\lambda_i}}))
\times e^{\mu n_xn_y\beta}}
\over {
|\prod_{\lambda_i}
(e^{\mu_0\beta}+e^{-\mu_0\beta}+\lambda_i+{1\over {\lambda_i}})
|e^{\mu_0n_xn_y\beta}}}=\cr
&e^{\mu n_xn_y\beta}
\sum_{n=0}^{n_xn_y}a_n(e^{\mu\beta}+e^{-\mu\beta})^n\cr}
\end{equation}
and the second on:
\begin{equation}\label{secondz}
\eqalign{
\tilde z(\mu) =&
{{(\prod_{\lambda_i}
(e^{-\mu\beta}+\lambda_i)(e^{-\mu\beta}+{1\over {\lambda_i}}))
\times e^{2\mu n_xn_y\beta}}
\over {
|\prod_{\lambda_i}
(e^{-\mu_0\beta}+\lambda_i)(e^{-\mu_0\beta}+{1\over {\lambda_i}})
|e^{2\mu_0n_xn_y\beta}}}=\cr
&e^{\mu n_xn_y\beta}
\sum_{n=-n_xn_y}^{n_xn_y}b_ne^{n\mu\beta}.\cr}
\end{equation}
where the $\lambda_i$ are the
eigenvalues of the matrix $P^-_{|_{\mu=0}}$.
Note that the expansion coefficients $b_n$ are the canonical
partition functions of the $n$-electron excitations
above half-filling, (with $n<0$ corresponding to holes).
We note here that eqns.(\ref{firstz}),(\ref{secondz}) follow
from the fact that the eigenvalues of $P^-_{|_{\mu=0}}$ are
either real or appear in complex conjugate pairs. The coefficients of
the characteristic polynomials, namely $\{a_n\}$ and $\{b_n\}$,
are obtained from (\ref{firstz}),(\ref{secondz}) by the
recursion procedure described in [\ref{BarbourBell}]. The sign problem
manifests itself in the fluctuating signs of these coefficients
from configuration to configuration of equilibrated Ising fields.
The expansion coefficients for a grand canonical
partition function (GCPF) are then
obtained
by averaging the coefficients of each of these polynomials over
the ensemble of configurations.
A similar procedure has been applied
in the study of the chiral phase transition in finite density lattice
QCD and in the evaluation of the critical mass
in lattice QCD [\ref{BarbourBell}].
\par
At $\mu_0=0$ (and at any imaginary chemical potential
[\ref{Dagotto}])
$\tilde z(\mu_0)$ is clearly positive.
With alternative choice of the updating $\mu_0$
the GCPF, $Z(\mu_0)$, is equal to the average sign of the weight
function $\tilde z(\mu_0)$[\ref{Morgen}],[\ref{Sugar?}].
We have performed calculations using
updating at half-filling and at $\mu_0\neq 0$.
The latter choice of the updating chemical
potential is important for simulations performed at low temperatures.
We will show below that it provides results consistent
with the half-filling updating at relatively high temperatures
($\beta\leq 5.$) while at higher values of $\beta $ it provides
better numerical stability in obtaining the expansion
coefficients of the GCPF, corresponding to
the finite hole occupation.
\par As the temperature is lowered the bounds on the eigenvalues of
the matrix $P^-_{|_{\mu=0}}$, which are found
via the Lanczos algorithm,
become very large.
Simulating
configurations at
$\beta=10.$, $dt=0.125$ we need to handle a lattice with
$n_\tau = 80$. For this set of parameters we find that the
eigenvalues vary between $10^{23}$ and
$10^{-23}$ on a $10^2$ lattice
which
damages severely the efficiency and the accuracy of the whole
calculation.
However, the characteristic
polynomial can be also obtained from the determinant
of the
the $(n_\tau n_xn_y)\times
(n_\tau n_xn_y) $
matrix $I+A^\pm$ given in (\ref{bigmatrix}), using the
eigenvalues of
$A^-_{|_{\mu=0}}$.
\par It follows from the structure of
$A^-_{|_{\mu=0}}$
that its eigenvalues
have a $Z_{n_\tau}$ symmetry:
if $\Lambda_i$ is an eigenvalue so is
$\Lambda_i e^{{{2\pi i n}\over {n_\tau}}}$ ($n=1,... n_\tau-1$).
The coefficients of the $\tilde z$ expansion in the fugacity powers
are actually functions of
the $\Lambda_i^{n_\tau}$.
The variation in these eigenvalues is significantly smaller, but
the matrix to be diagonalized is $n_\tau ^2$ times bigger, leading
to a more time consuming diagonalization procedure.
\par The method used in our calculations consists of
representing the determinant of (\ref{bigmatrix}) as
\begin{equation}\label{midmatrix}
det(I+\cal{A})
\end{equation}
with
\begin{equation}
{\cal{A}=\left(\matrix{0&0& ...&B_1 B_2 ... B_{n_t}\cr
-B_{n_t+1} B_{n_t+2}....&0&...&0\cr
...&...&...&... \cr
...&...&...&... \cr
0&0&-...B_{n_\tau-1}B_{n_\tau}&0\cr}\right)}.
\end{equation}
where the matrix $\cal{A}$ is of the size
$(n_xn_y{{n_\tau}\over {n_t}})\times(n_xn_y{{n_\tau}\over {n_t}})$.
(In the last equation the $B$ matrices are taken at $\mu=0$.)
The eigenvalues of this matrix have a reduced
symmetry $Z_{n_{\tau}}\rightarrow Z_{n_\tau /n_t}$ and thus
variations of a larger magnitude than those of $A$, but
the diagonalization procedure is more
efficient.
On the other hand its eigenvalues are varying in a smaller
range than the eigenvalues of the total time ordered
product $\prod_{l=1}^{n_\tau}B_l $.
By finding the most appropriate ratio ${n_\tau /n_t}$ we
succeed to obtain the eigenvalues of ${\cal A}$ with the
required precision and then taking the ${n_\tau /n_t}$
power of them we obtain the $\{\lambda_i\}$ and evaluate
the expansions (\ref{firstz}) and (\ref{secondz}).
Note that
since the coefficients of the characteristic polynomials for
(\ref{firstz})
and (\ref{secondz}) depend only on $\lambda_i+
\lambda_i^{-1}$, the above procedure, although introducing large
variations in the $\lambda_i$'s themselves, does not lead
to significant errors in the coefficients around
half-filling. It is these coefficients which
determine the behaviour of the smallest zeros
of the GCPF and thus the phase structure of the model.
\par The physical observables are derived using
eq.(\ref{observb}).
For a given configuration of Ising fields $\{s_{i,l}\}$,
the value of an operator $O$
can be calculated as a polynomial in the fugacity variable.
Knowing the coefficient of each power
of the fugacity in the $\tilde z$ expansion, one can then
easily construct the corresponding coefficients for the contribution
of this term to the observable by averaging each coefficient
over the equilibrated ensemble.
In this paper we measure only the expansion of the GCPF and hence
we can predict the critical value
of the chemical potential for which the relative weight of the
the state with a finite doping fraction will be of the same
order as the half-filled state. We expect that this prediction
will be reflected in the behaviour of the hole occupation number
and consistent with
the critical values of $\mu$ obtained from an analysis of
the complex zeros of the GCPF. Moreover, following
the suggestion of Dagotto et al.[\ref{Dagotto}] we evaluate the
energy gap between the one pair state and the half-filled
state.
\bigskip\newline\noindent{\bf 3. Simulations and measurements}
\bigskip\newline\noindent
The simulation procedure is based on the algorithm described by
White {\it et al.} [\ref{White}]. We use Metropolis algorithm for
updating the configurations of Ising variables. Here
we describe the procedure for
the half-filling updating. (For clarity,
we omit the spin labels in the following.)
The generalization for the
finite $\mu_0 $ updating is straightforward.
\par The simulation starts from an arbitrary configuration for which
we calculate the equal-time Green's function on a time slice $l$
\begin{equation}\label{propagator}
G(l)=(I+P(l))^{-1}
\end{equation}
where $P(l)$ is a time ordered product of the form
\begin{equation}\label{string}
B_l...B_1B_{n_\tau}...B_{l+1}
\end{equation}
To reduce the numerical errors in the evaluation of these
products we apply the modified Gram-Schmidt (MGS) decomposition
as proposed by
White {\it et al.} [\ref{White}].
Decomposing
the products of each four matrices in
(\ref{string})
into a product of an orthogonal matrix,
a diagonal one and an upper-triangular matrix whose diagonal
elements equal to one
enables us
to deal with large variations in the matrix elements.
The inversion of the $(I+P(l))$ appearing
in the r.h.s. of (\ref{propagator}) is also simplified by
the MGS procedure.
Once the equal-time propagator is evaluated
on a time-slice $l$ we flip one of the spins $s_{i,l}$ and
and accept the new configuration with respect to the ratio of the
new and old statistical weights, defined as:
\begin{equation}\label{ratio}
{{det M'^2}\over {det M^2}}\times e^{\lambda\delta(s_{i,l})}
\end{equation}
where
$\delta(s_{i,l})$ is the difference in the potential term
due to the flipped spin.
\par This ratio is determined by the
value of the equal-time Green's function
(by its diagonal term corresponding to the flipped spin) and by
a matrix $\Delta$
with only one nonzero element:
\begin{equation}
\Delta(i,l)_{j,k} = e^{-2dt\lambda s_{i,l}}\delta_{i,j}\delta_{j,k}
\end{equation}
If the new configuration is accepted we
calculate the new Green's function
$G(l)'$ corresponding to this configuration using:
\begin{equation}\label{etupd}
G(l)'=G(l)-{1\over R}\times G(l)\Delta(i,l)(I-G(l))
\end{equation}
with
\begin{equation}\nonumber
R={{det M'}\over {det M}}=1+(1-G_{ii})\Delta(i,l)_{ii}
\end{equation}
The nonlocal impact of the updated configuration is represented
in the new Green's function by eq. (\ref{etupd}).
\par When the updating of the Ising fields on the l-th time-slice is
completed we move to the next time slice using the relation:
\begin{equation}\label{prop}
G(l+1)=B_{l+1} G(l) B_{l+1}^{-1}.
\end{equation}
Following the suggestion of
[\ref{White}], we evaluated the Green's function
from scratch every four time steps. This procedure is very
time consuming, but is required in order that the numerical
errors accumulated using eq.(\ref{prop}) are kept under control.
Since
the hopping term in the Hamiltonian is constant,
the construction
of the Green's function from scratch reduces to a redefinition
of the interaction matrix $e^{dt V}$ on each time-slice due to
the updated Ising fields.
The computational effort involved in this procedure is minor as
$V$ is a diagonal matrix
and its exponentiation is fast. The hopping part is exponentiated
only once at the start of the simulation procedure. Instead
of using the checkerboard decomposition suggested by White {\it et al.}
we expanded $e^{-dt K}$ taking advantage of its sparseness. Since
this expansion is performed only once it can be done up to an arbitrary
high order. We checked that taking a 10-th order
expansion provided us
with sufficiently high precision for the parameters used in the
simulations described below.
\bigskip
\newline\noindent{\bf 4. Results and conclusions}
\bigskip\newline\noindent
The expansion coefficients of the GCPF as a polynomial in
the fugacity variable were calculated at several values of the
inverse temperature $\beta $ with the Coulomb repulsion fixed at
$U=4t$
(see eq.(\ref{Hubbard}).
We present results
obtained
from simulations performed at half-filling and at
chemical potential $\mu_0=0.9$. The spatial size of the lattice
is $4^2$ throughout apart from one simulation at half-filling on a
$10^2$ lattice at $\beta=10.0$.
\par {\bf Half-filling results:}
The half-filling simulation was performed at $\beta=0.3,1.2,
2.5,5.0$ and $10.0$
with $n_\tau=4,16,20,40$ and $80$ respectively.
The
euclidean time-spacing $dt$ was varied to
check the numerical stability of the simulations and to
allow comparison with the results of other groups [\ref{Moreo}].
The number of configurations required to get sufficiently low
errors in the expansion coefficients is, in general, greater than 2000
and increases with $\beta$.
The particle-hole symmetry is manifested via the equality between the
coefficients of $e^{\mu n\beta}$ and $e^{-\mu n\beta}$.
This symmetry follows directly from eqs.(\ref{firstz},\ref{secondz}).
The $z_n=<b_n>$ coefficient corresponds to the contribution
of the $n$-hole
state to the GCPF (canonical parition functions)[\ref{Dagotto}]
and $z_0$ is the canonical partition function for the
half-filled state. The latter is obtained with
low error after a small number of measurements. The higher order
coefficients require averaging over a larger number of configurations.
For $\beta \leq 5$ all the $2(n_xn_y)^2+1$
averaged coefficients were found to be positive.
\par As we extend our treatment to larger values of $\beta$,
negative
coefficients appear in the expansion of the GCPF, but with
large errors.
Note that the coefficients arising from
a single configuration
do not have to be positive, as only the ensemble averages
are identified as the canonical partition function at a given
particle number.
However, the low temperature simulation at half-filling does
lead to a high
variation of the coefficients in different field configurations
and thus to the high errors. This is because updating at $\mu=0$
minimizes the fluctuations in only $z_0$ (half-filling) which
dominates the statistical weight at high $\beta$.
The large variations in these coefficients indicates that this
statistical weight, namely the determinant
at $\mu_0=0$, becomes inefficient as
$\beta$ gets large.
A more appropriate
choice is to update at $\mu_0 \neq 0$.
\par {\bf Simulations at $\mu_0 \neq 0$:}
We performed our simulations at $\mu_0=0.9$
and updated with respect to the absolute value of the weight
function as described in the previous section at
$\beta=2.5,3.,5.0,5.4,6.0$ and $7.5$ with
$n_\tau=20,40,40,72,48$ and $60$ respectively.
In this simulation the value of the GCPF at $\mu_0$ is the
average sign of $det(M^+)det(M^-)$. This requirement provides a
useful check as to our numerical accuracy in extracting the
coefficients of the characteristic polynomial.
At $\beta=2.5$ and $\beta=5.$ we compared the results
of these simulations
with those performed at half-filling updating and found that
the normalized coefficients
are equal within the statistical error.
\par In Table 1 and Fig.1
we present the expansion coefficients based on eq. (\ref{secondz})
obtained
from simulations performed either at half-filling or at updating
chemical potential $\mu_0=0.9$.
The normalization of the GCPF is chosen such that $Z(\mu=0)=1.$
Our results are
consistent with those obtained by Moreo {\it et al.}[\ref{Moreo}].
\par Fig.1 shows the coefficients
$z_n$ ($n=0,1,2,3$) as a function of $\beta$,
with normalization such that the GCPF $Z(\mu =0)=1$.
The general
tendency is a sharp decrease of the coefficients with
higher occupation number.
\par Figs.2 and 3 show the hole density
and the susceptibility respectively,
as a function of the chemical potential.
In these exploratory simulations, the coefficients
corresponding to high occupation number are
determined with large errors.
However,the peak in the
susceptibility at $\mu \approx 1.0$ is due to
the low occupation levels which are determined with small errors.
The structures at
$\mu > 1.2$ do depend on the
levels with large errors and require further
investigation.
As the temperature decreases the peak in the susceptibility
sharpens significantly
in the region $0.75<\mu<1.25$,
especially for $\beta >5.0$. At that $\beta$
the susceptibility does seem to signal some change in behaviour.
This could be associated with the onset of
a phase transition related to the occupation
of holes. We explore this possibility further by analyzing
the zeros of the GCPF
in the complex fugacity plane.
\par We do this by finding the zeros of the averaged polynomial
eq.(\ref{firstz}).
Since the large $n$ coefficients are evaluated with relatively low
precision we checked the stability of the small zeros under truncation
of the polynomial to $n=4$.
Table 2 gives the two smallest zeros for various values of
the inverse temperature.
In Fig.4 we plot these zeros of the partition
function in the first quadrant of the
complex $\mu$ plane. The zeros have the symmetries that if $\mu$
is a zero, so also is $-\mu$ and their complex conjugates.
\par To a very good approximation, the imaginary part of these
zeros of the partition function scales as
${\pi\over \beta}$.
For any finite value of $\beta$
the fugacity would remain negative yielding no phase transition
in the physical region. However, at zero temperature ($\beta=\infty$)
the Im$(\mu_c)$ vanishes and a phase transition may be possible.
To extrapolate to the low temperature behaviour
of the zeros we note an almost linear dependence
between the imaginary part of the lowest zero
and its real part, for values
of $\beta \geq 5.$ The linear fit of these zeros is shown
in Fig.5. To clarify this point we show in Fig.6 the linear
fit of  Re$(\mu)\times \beta$ {\it vs.} $\beta$ in the same
region of $\beta$. The lowest zero at
$\beta= 2.5$ and 3 does not exhibit this scaling behaviour.
Measuring the lowest zero of the partition function
at high temperatures ($\beta=0.3$ and $\beta=1.2$)
we find that the real parts of the zero to be
3.0 and
0.721 respectively with imaginary part ${\pi \over \beta}$.
The above is consistent with the lowest zero, $\mu_c$,
scaling such that
Re$(\mu_c)\times\beta$ is constant in the high
temperature regime. At lower temperatures,
$\beta=2.5-3$, there is a crossover region into the low temperature
regime where
\begin{equation}\label{scaling}
Re\mu_c=-2.5/\beta+1.1.
\end{equation}
This qualitative distinction between the high and low temperature
regimes arises from a clear difference in the relative
contributions of the finite particle number states to the grand
canonical partition function (see Table 1).
In the high temperature regime the first two canonical partition
functions are of the same order as the half-filled level and hence
the corresponding states are excited even at zero chemical
potential. On the other hand, these states in the low temperature GCPF
are only excited by non-zero chemical potential.
Based on the above observations, we conclude that there is the
possibility of a phase transition at zero temperature and
$\mu_c \approx 1.1$
but that
no signal has been found for a phase transition at $T>0$.
Of course, the above simulations have been performed on a small system.
There may well be large finite volume effects.
\par
The above conclusion - that there may be a phase transition at zero
temperature - will not alter if the lowest zero in the fugacity
plane remains real (and hence necessarily negative). If there is a
phase transition at nonzero temperature, then some complex zeros
must be in the complex fugacity plane with $Re(e^{\mu \beta})>0$ and
pinch the positive real axis in the infinite spatial volume limit. No
signal of the possible onset of this mechanism was observed on the
$4^2$ lattice, {\it i.e.} no zeros were found in the first or fourth
quadrants of the complex $e^{\mu \beta}$ plane.
\par
As we increase the spatial lattice size we should therefore observe
either an increase in the density of zeros adjacent to, or on,
the negative real fugacity axis or, if the
alternative mechanism is masked by finite size effects on the $4^2$
lattice, a migration of zeros to the $Re(e^{\mu \beta})>0$ complex
half-plane. It is also important to confirm the scaling behaviour
of eq.({\ref{scaling})
for the lowest zero. Zeros adjacent to this one should
also scale in an analogous manner with the temperature so that there
is a well defined locus of zeros in the zero temperature limit.
We intend to extend our simulations to $6^2$ and $8^2$ spatial
lattices.
\par The absence of a critical positive fugacity above zero temperature
can be interpreted in part as a realization of the Mermin-Wagner theorem
[\ref{MerminW}], which claims absence of magnetic ordering in two
dimensional spin systems at non-zero temperature.
This theorem is in particular relevant to HTSC
models based on the Heisenberg antiferromagnetic Hamiltonian
resulting in the strong coupling treatment of the Hubbard model.
If the isotropy of Heisenberg antiferromagnet is violated, {\it e.g.}
by an interlayer interaction, the conditions of the theorem do not hold,
thus relating the vanishing critical temperature for the
superconductivity to the
isotropy of the effective nearest neighbours coupling[\ref{Huang}].
The 2D Hubbard model with effective interlayer interaction was recently
studied by M. Frick {\it et al.}
using the
Projector Monte Carlo technique[\ref{Frick}],[\ref{Frick1}]. Their
results show
some evidence for HTSC.
We note that the analysis described in our work
can also be applied to
the extended Hubbard model including effective interlayer interactions.
\par Finally, following the suggestion of Dagotto {\it et al.}
[\ref{Dagotto}]
we calculate the binding energy of holes.
The energy gap between the two and one-hole ground states, $E_2-E_1$,
is given by the slope of the linear fit to the ratio
$\log{{z_2}\over {z_1}}$ {\it vs.} $\beta$
(see Figs.7,8). Fig.7 presents the fit
for the data in the range $2.5\leq\beta\leq7.5$
while, in Fig.8, we fit only the results for $\beta\geq 5.0$
in the light of the discussion above.
The first fit shows an energy gap of $0.87\pm0.02$ which is very close
to the result of Dagotto {\it et al.}
($0.88\pm.0.02)$[\ref{footn}].
The energy gap obtained
from the second fit is $0.85\pm0.02$. Analogous fits, (Figs.9,10), for
$\log{{z_1}\over {z_0}}$
{\it vs.} $\beta$ give the energy gap between the
ground state with one hole and the corresponding state
at half-filling, $E_1-E_0$.
The fits give $E_1-E_0=
0.83\pm0.02$ and $0.95\pm0.02$ respectively.
The binding energy of holes is given by the difference
\begin{equation}\label{gap}
(E_2-E_1)-(E_1-E_0).
\end{equation}
Thus the first fit yields positive binding energy with no pair
creation expected while the latter suggsts a
binding energy of $-0.1$.
Since this derivation of the energy gaps is valid only in the
low temperature regime, which is distinct
from the high temperature one, the second fit seems to be more
appropriate. We note that we have not included the spin wave
contribution to the ratio of the canonical partition functions
$z_1$ and $z_0$[\ref{Dagotto}], since at low temperature, the
spin wave contribution should become small.
The result of our low temperature
fit is close to that obtained by Dagotto {\it et al.}
$(E_1-E_0=0.98\pm0.02)$. However, their
fit included lower $\beta$ data thus
requiring a spin wave contribution.
Inclusion of a spin wave contribution to our
canonical partition function at half-filling
would raise the estimate of the one-hole
ground state energy even higher, thus increasing the binding energy.
\par We conclude with suggested
extensions of the above method. The predicted
zero temperature phase transition can be confirmed by a lower
temperature study. With this in mind,
simulations at $\beta=12$ ($n_\tau=160$)
are under current investigation.
A study of the finite size effects, in particular, the
scaling properties of Im$(\mu_c)$ as a function of the volume
is also necessary. It is also important to perform longer
simulations in order that the analysis can be extended to
larger values of the chemical potential.
One can also
generalize the method described above to derive polynomial
expansions in the fugacity variable for
other physical observables and thus extend the study
of the nature of the phase transition and of
its possible relevance
to high T$_c$ superconductivity. This study would involve
examining the persistence of the antiferromagnetic order
into the finite doping region[\ref{Lilly}].
\bigskip
\newline\noindent{\bf Acknowledgements}
\bigskip\noindent\newline
We thank A. Moreo, D.J. Scalapino, D.G. Sutherland, S. Hands
and A. Kovner for illuminating discussions in the course
of this work. We also are grateful to C.T.H. Davies for her
constructive comments.
\vfil
\newpage
{\bf \noindent References}
\begin{enumerate}
\item\label{Hirsch}
{J.E. Hirsch, {\it Phys.Rev.} {\bf B31}, 4403 (1985).}
\item\label{Blankenbecler}
{R. Blankenbecler, D.J. Scalapino and R.L. Sugar,
{\it Phys.Rev.} {\bf D40}, 2278 (1981).}
\item\label{Dagotto}
{E. Dagotto et al.
, Santa Barbara preprint NSF-ITP-89-137;\\
A. Moreo, {\it Nucl. Phys.}{\bf B17}, (Proc.
Suppl.), 716 (1990).}
\item\label{White}
{S.R. White {\it et al.}, {\it Phys.Rev.} {\bf B40}, 506 (1989).}
\item\label{bookH}
{A.P. Balachandran, E. Ercolessi, G. Morandi and A.M. Srivastava,
{\bf Hubbard Model and Anyon Superconductivity} {\it World
Scientific 1990}.}
\item\label{Anderson}
{P.W. Anderson,
{\it Science} {\bf 235} , 1196 (1987).}
\item\label{YangLee}
{C.N. Yang and T.D. Lee, {\it Phys. Rev.} {\bf 87}, 404 (1952);
\newline\noindent
T.D. Lee and C.N. Yang, {\it Phys. Rev.} {\bf 87}, 410 (1952).}
\item\label{BarbourZabeur}
{I.M. Barbour and Z.A. Sabeur, {\it Nucl. Phys.}{\bf B342},
269 (1990).}
\item\label{Suzuki}
{M. Suzuki, {\it Prog. Theor. Phys.} {\bf 56}, 1454 (1976).}
\item\label{SugarLat}
{R.L. Sugar, {\it Nucl. Phys.} {\bf B17} (Proc. Suppl.), 39 (1990).}
\item\label{BarbourBell}
{I.M. Barbour and A.J. Bell,
{\it Nucl. Phys.}{\bf B372}, 385 (1992).}
\item\label{Morgen}
{W. von der Linden, I Morgenstern and H. de Raedt,
{\it Phys. Rev.} {\bf B41}, 4669 (1990).}
\item\label{Sugar?}
{E.Y. Loh Jr.{\it et al.} {\it Phys. Rev.} {\bf B41}, 9301 (1990).}
\item\label{Moreo}
{A. Moreo, Private communication.}
\item\label{MerminW}
{N.D. Mermin and H. Wagner, {\it Phys. Rev. Lett} {\bf 17}, 1133
(1966).}
\item\label{Huang}
{K. Huang and E. Manousakis, {\it Phys. Rev.} {\bf B36}, 8302
(1987).}
\item\label{Frick}
{M. Frick, I. Morgenstern and W. von der Linden, {\it Int. Jour. of Mod.
Phys.}{\bf C3}, 102 (1992).}
\item\label{Frick1}
{M. Frick, I. Morgenstern and W. von der Linden, {\it Z. Phys}
{\bf B82}, 339 (1991).}
\item\label{Lilly}
{L. Lilly, A. Muramatsu and W. Hanke, {\it Phys. Rev. Lett.}
{\bf 65}, 1379 (1990);\\
A. Krasnitz, A. Kovner and E.G. Klepfish, {\it Phys. Rev.}{\bf B39},
9147 (1989).}
\item\label{footn}
{All the energies are given
in units of the hopping parameter $t$- see Eq.(\ref{Hubbard}).}
\end{enumerate}
\newpage
\bigskip
\begin{center}
{\bf Table 1.}\\
\bigskip
\begin{tabular}{|c|c|c|c|c|c|c|c|c|}
\hline
   & \multicolumn{2}{c|}{$z_0$} & \multicolumn{2}{c|}{$z_1$}
& \multicolumn{2}{c|}{$z_2$}
&\multicolumn{2}{c|}{$z_3$} \\
\hline
$\beta$: & coeff. & err. & coeff. & err.
& coeff. & err.
& coeff. & err. \\
\hline
0.3 & 0.171 & 2.E-4 & 0.156 & 1.5E-4 &
0.119 & 3.E-4 &
0.074 & 6.E-5 \\
\hline
1.2 & 0.297 & 0.001 & 0.223 & 2.E-4 &
0.097 & 4.E-4 &
0.025 & 2.5E-4 \\
\hline
2.5 & 0.477 & 0.002 & 0.220 & 0.001 &
0.039 & 5.E-4 &
0.003 & 8.E-5 \\
\hline
3. & 0.575 & 0.016 & 0.189 & 0.002
& 0.022 & 1.E-4
& 0.001 & 1.E-5 \\
\hline
5. & 0.872 & 0.005 & 0.063 & 0.002
&0.001 & 2.E-4
& 5.E-6 & 3.E-6 \\
\hline
5.4 & 0.901 & 0.061 & 0.049 & 0.002
& 7.E-4 & 3.E-5
& 4.E-6 & 2.E-7 \\
\hline
6. & 0.946 & 0.056 & 0.027 & 0.001
& 2.E-3 & 1.E-5
& 6.E-7 & 4.E-8 \\
\hline
7.5 & 0.986 & 0.063 & 0.007 & 5.E-4
& 2.E-5 & 1.E-6
& 1.E-8 & 1.E-9 \\
\hline
\end{tabular}
\end{center}
\newpage
\bigskip
\begin{center}
{\bf Table 2.}\\
\bigskip
\begin{tabular}{|c|c|c|c|c|c|c|}
\hline
   & \multicolumn{2}{c|}{Re$\mu$} & \multicolumn{2}{c|}{Im$\mu$}
& \multicolumn{2}{c|}{(Im$\mu )\times\beta$} \\
\hline
$\beta$: & Full & Trunc. & Full & Trunc.
& Full & Trunc. \\
\hline
2.5 & 0.526 & 0.464 & 1.257 & 1.257 &
3.141 & 3.141 \\
    & 0.821 & 0.731 & 1.257 & 0.952 &
3.141 & 2.380 \\
\hline
3. & 0.582 & 0538 & 1.047 & 1.047 &
 3.141 & 3.141 \\
   & 0.764 & 0.749 & 1.047 & 0.819 &
 3.141 & 2.457 \\
\hline
5. & 0.609 & 0.607 & 0.628 & 0.628 &
 3.140 & 3.140 \\
   & 0.775 & 0.818 & 0.628 & 0.628 &
 3.140 & 3.140 \\
\hline
5.4 & 0.627 & 0.625 & 0.582 & 0.582
& 3.143 & 3.143 \\
    & 0.884 & 0.820 & 0.572 & 0.477
& 3.089 & 2.576 \\
\hline
6. & 0.685 & 0.680 & 0.524 & 0.524
& 3.144 & 3.144 \\
   & 0.801 & 0.848 & 0.524 & 0.453
& 3.144 & 2.718 \\
\hline
7.5 & 0.768 & 0.785 & 0.380 & 0.397
& 2.850 & 2.978 \\
    & 0.990 & 0.872 & 0.310 & 0.419
& 2.325 & 3.142 \\
\hline
\end{tabular}
\end{center}
\newpage
{\bf \noindent Figure Captions}
\begin{enumerate}
\item
{Fig.1 $z_0,z_1,z_2$, and $z_3$ as a function of $\beta$. Their
associated statistical errors are also shown if larger
than the resolution.}
\item
{Fig.2 Average hole density (doping fraction)
as a function of $\mu$ for $\beta$
between 2.5 and 7.5:\\
$\beta=2.5$ and $3$ --- solid lines\\
$\beta=5.0$ --- dashed line\\
$\beta=5.4$ --- dashed-dotted line\\
$\beta=6.0$ --- dotted line\\
$\beta=7.5$ --- solid line;}
\item
{Fig.3 Susceptibility as a function of $\mu$ for $\beta$
between 2.5 and 7.5:\\
$\beta=2.5$ and $3$ --- solid lines\\
$\beta=5.0$ --- dashed line\\
$\beta=5.4$ --- dashed-dotted line\\
$\beta=6.0$ --- dotted line\\
$\beta=7.5$ --- solid line;}
\item
{Fig.4 Two smallest zeros of the GCPF in the first quadrant of the
complex $\mu$ plane
for $\beta$ between 2.5 and 7.5:\\
$\beta=2.5$ --- white dotted squares\\
$\beta=3.0$ --- black crosses\\
$\beta=5.0$ --- white circles\\
$\beta=5.4$ --- black circles\\
$\beta=6.0$ --- black squares\\
$\beta=7.5$ --- white squares;}
\item
{Fig.5
Im$(\mu_c)$ plotted against Re$(\mu_c)$. The solid line is a
linear fit for $\beta$ between 5.0 and 7.5;}
\item
{Fig.6
Re$(\mu_c)\beta$ as a function of $\beta$. The solid line is a
linear fit for $beta$ between 5.0 and 7.5;}
\item
{Fig.7 $\log(z_2 / z_1)$ as a function of $\beta$. The solid line
is a linear fit for $2.5\leq\beta\leq7.5$;}
\item
{Fig.8 $\log(z_2 / z_1)$ as a function of $\beta$. The solid line
is a linear fit for $5.0\leq\beta\leq7.5$;}
\item
{Fig.9 $\log(z_1 / z_0)$ as a function of $\beta$. The solid line
is a linear fit for $2.5\leq\beta\leq7.5$;}
\item
{Fig.10 $\log(z_1 / z_0)$ as a function of $\beta$. The solid line
is a linear fit for $5.0\leq\beta\leq7.5$;}
\end{enumerate}
\newpage
{\bf \noindent Table Captions}
\begin{enumerate}
\item
{Table 1
Expansion coefficients of the GCPF for $\beta$ between
0.3 and 2.5.
z$_0$ corresponds to
the half-filling and z$_n$ to the $n-$hole coefficients.}
\item
{Table 2
Smallest zeros of the GCPF in the complex $\mu$ plane for
$\beta$
between 2.5 and 7.5 .
The zeros are obtained from the full polynomial and the truncated
one (up to 4-hole coefficient), to check the numerical stability
of the results.}
\end{enumerate}
\end{document}